\documentclass[prl,twocolumn,superscriptaddress,amsmath,english,preprintnumbers,amssymb,showpacs]{revtex4}
\usepackage[T1]{fontenc}
\usepackage[latin1]{inputenc}
\usepackage{graphicx}
\newcommand{\beq}{\begin{equation}}
\newcommand{\eneq}{\end{equation}}
\newcommand{\beqy}{\begin{eqnarray}}
\newcommand{\eneqy}{\end{eqnarray}}
\makeatletter
\usepackage{amsmath}
\usepackage{epsfig}
\usepackage{dcolumn}
\usepackage{bm}
\usepackage{rotating}

\input{epsf}
\usepackage{color}
\usepackage{graphicx}
\usepackage{mathrsfs}
\usepackage{theorem}
\usepackage{amsmath,amssymb}

\newcommand{\ket}[1]{\left| #1 \right\rangle}
\newcommand{\bra}[1]{\left\langle #1 \right|}

\usepackage{dcolumn}
\usepackage{bm}
\usepackage{times}
\usepackage{multirow}

\def\bra#1{\mathinner{\langle{#1}|}}
\def\ket#1{\mathinner{|{#1}\rangle}}

\usepackage{babel}
\makeatother
\begin{document}

\title{Perfect mirror transport protocol with higher dimensional quantum chains}
\author {Gerardo~A. Paz-Silva}
\affiliation{Centre for Quantum Computer Technology, Macquarie University, Sydney, NSW 2109, Australia}
\author {Stojan Rebi\'{c}}\author{Jason Twamley}
\affiliation{Centre for Quantum Computer Technology, Macquarie University, Sydney, NSW 2109, Australia}
\author {Tim Duty}
\affiliation{Department of Physics, University of Queensland, St Lucia, QLD 4072, Australia}

\begin{abstract}
A globally controlled scheme for quantum transport is proposed. The scheme works on a 1D chain of nearest neighbor coupled systems of qudits (finite dimension), or qunats (continuous variable), taking any arbitrary initial quantum state of the chain and producing a final quantum state which is  perfectly spatially mirrored about the mid-point of the chain. As a particular novel application, the method can be used to transport continuous variable quantum states. A physical realization is proposed where it is shown how the quantum states of the microwave fields held in a chain of driven superconducting coplanar waveguides can experience quantum mirror transport when coupled by switchable Cooper Pair Boxes.
\end{abstract}

\maketitle
\indent 
A fundamental issue in quantum information theory and the development of a quantum computers is the transfer of a quantum state through a reliable, high fidelity, method. Such methods are known as {\it quantum wires}, or q-wires. Q-wire design should avoid the detailed control of its components as this may introduce additional errors which corrupt the transport and typically implies challenges in implementation. Ideally such quantum transport should be capable of transporting entire entangled registers of quantum information with near perfect fidelity and operate with the minimum of external control. Recently interest in q-wires has grown considerably due to the advent of higher dimensional quantum systems (qudits or CV, continuous variables) as components in quantum processors.

So far, the only proposals for a mirror transport involved qubit states \cite{Bose:2003p3539} and generally required significant resources to achieve useful transport fidelity \cite{Christandl:2004p3626,Burgarth:2005p3628}. In this Letter a new protocol is described that for the first time formulates and solves the problem of quantum mirroring in chains consisting of nearest neighbor coupled identical qu{\bf d}it systems or  infinite dimensional qu{\bf n}at [Continuous Variable (CV)] systems. It is based on a cellular automata-like scheme for perfect mirror transport~\cite{Fitzsimons:2006p852}, spatially inverting a quantum state distributed along a chain via the application of homogenous single qubit pulses and Ising-like nearest neighbor interactions. 

Qudits and qunats often lead to features which have certain advantages over qubits~\cite{BechmannPasquinucci:2000p3575,MolinaTerriza:2005p3622}. Fault tolerant quantum computation has been shown for qudits \cite{Gottesman:1999p3585}, and schemes for trapped ion qudit quantum computation have been developed \cite{Klimov:2003p3618}. One-way quantum computing via qudit cluster states is possible, as is the formulation of qudit decoherence free subspaces and topologically protected qudit memories \cite{bishop:012314}. Moreover, continuous variable (CV) quantum computation has been widely studied  in recent years \cite{Braunstein:2005p3364}, while CV quantum transport has also attracted recent interest \cite{Plenio:2004p3655}. Our CV results below provide an alternative method for perfect quantum transport in CV chains. 

The aim is to develop a quantum mirror transport protocol for qudit and CV systems. The qubit case was considered previously ~\cite{Fitzsimons:2006p852}. In this Letter a scheme suitable for qudits and CVs is developed. Physical realization is proposed involving perfect spatial mirroring of continuous variables in Circuit-QED \cite{Blais:2004p3701}, spatially mirroring a quantum state of microwave photons trapped in a chain of coplanar waveguides (CPWs), coupled via Cooper pair boxes (CPBs). 

{\em Qudit mirror transport.} It will prove useful to recall some basic properties of qudit quantum gates \cite{Daboul:2003p3415,Schwinger:1960p5256}. A qudit is a state of a $d-$level system described by a $d-$dimensional Hilbert space $\mathcal{H}_d$ with basis $\{|k\rangle;\, k=0,1,\cdots,d-1\}$. Generalized Pauli operators are defined via $\hat{X}_d\equiv \sum_{j=0}^{d-2}\, |j\oplus 1\rangle\langle j|$ and $\hat{Z}_d\equiv \sum_{j=0}^{d-1}\,|j\rangle \zeta_d^j \langle j|$, where $\zeta_d$ is a $d-$th root of unity $\zeta_d^d=1$, and $\oplus$ is addition modulo $d$.  As for qubits, the qudit basis is $\hat{\Xi}_d^{ij}\equiv \hat{X}_d^i\hat{Z}_d^j$, with the commutation relation $\hat{X}_d^j\hat{Z}_d^k = \zeta_d^{jk} \hat{Z}_d^k \hat{X}_k^j$ and $\hat{X}_d^d = \hat{Z}_d^d = \hat{\mathbb{I}}$. In the qudit case the Hadamard gate is replaced by a Fourier gate $\hat{F}$, given by $\hat{F} \ket{a} = d^{-1/2} \sum_k \zeta^{ k a} \ket{k}$, and satisfying $\hat{F}^4 = \hat{\mathbb{I}}$ and $\hat{F}^2 \ket{a} = \ket{-a}$ ($-a$ is taken modulo $d$). 

Basic gates needed for our protocol are the qudit analogues of the CNOT (i.e. SUM), CPHASE and SWAP gates. Suppressing the dimension $d$ and indicating the (control, target) qudits as subscripts \cite{Wang:2003p3414,Daboul:2003p3415},
\begin{subequations}
\label{Eqs:gates}
\begin{eqnarray}
\hat{U}_{CPHASE} &=& \sum^{d-1}_{n=0} \ket{n}\bra{n}_{(1)} \otimes \hat{Z}^n_{(2)} = \hat{S}_{(12)} \;\;,\label{Phase1} 
\end{eqnarray}
\begin{eqnarray}
\hat{U}_{SUM ({12})} &=& \sum^{d-1}_{n=0} \ket{n}\bra{n}_{(1)} \otimes \hat{X}_{(2)}^n= \hat{D}_{(12)} \;\;,\label{Sum1}\\
&=&\hat{F}^{-1}_{(2)} \hat{U}_{CPHASE({12})} \hat{F}_{(2)}\label{Sum2}\;\;, \\
\hat{U}_{SWAP} &=& \hat{D}_{(12)} \hat{F}^2_{(1)}\hat{D}_{(21)} \hat{F}^2_{(1)} \hat{D}_{(12)} \hat{F}^2_{(2)}\;\;,\label{Swap1}
\end{eqnarray}
\end{subequations}
$\hat{A}_{(y)}$ denoting the operator $\hat{A}$ acting on qudit $y$. Decompositions (\ref{Swap1}) and (\ref{Sum2}) (obtained using $\hat{Z}\hat{F}=\hat{F}\hat{X}$) are very similar in form to those found for the corresponding qubit gates: $SWAP=CNOT_{12}CNOT_{21}CNOT_{12}$ and $CNOT_{12}=(\hat{\mathbb{I}}\otimes \hat{H})CPHASE (\hat{\mathbb{I}}\otimes \hat{H})$. 
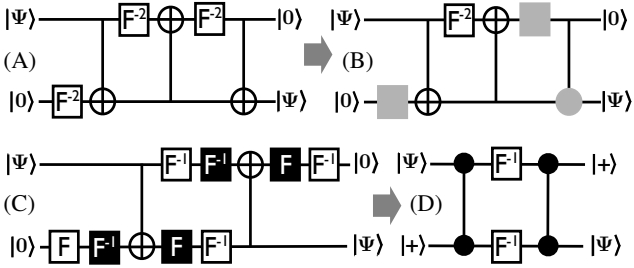
\begin{figure}[t]
\begin{center}
\setlength{\unitlength}{1cm}
\begin{picture}(4,3.3)
\put(-2.2,-3.1){\includegraphics[width=11.5cm,height=6.6cm]{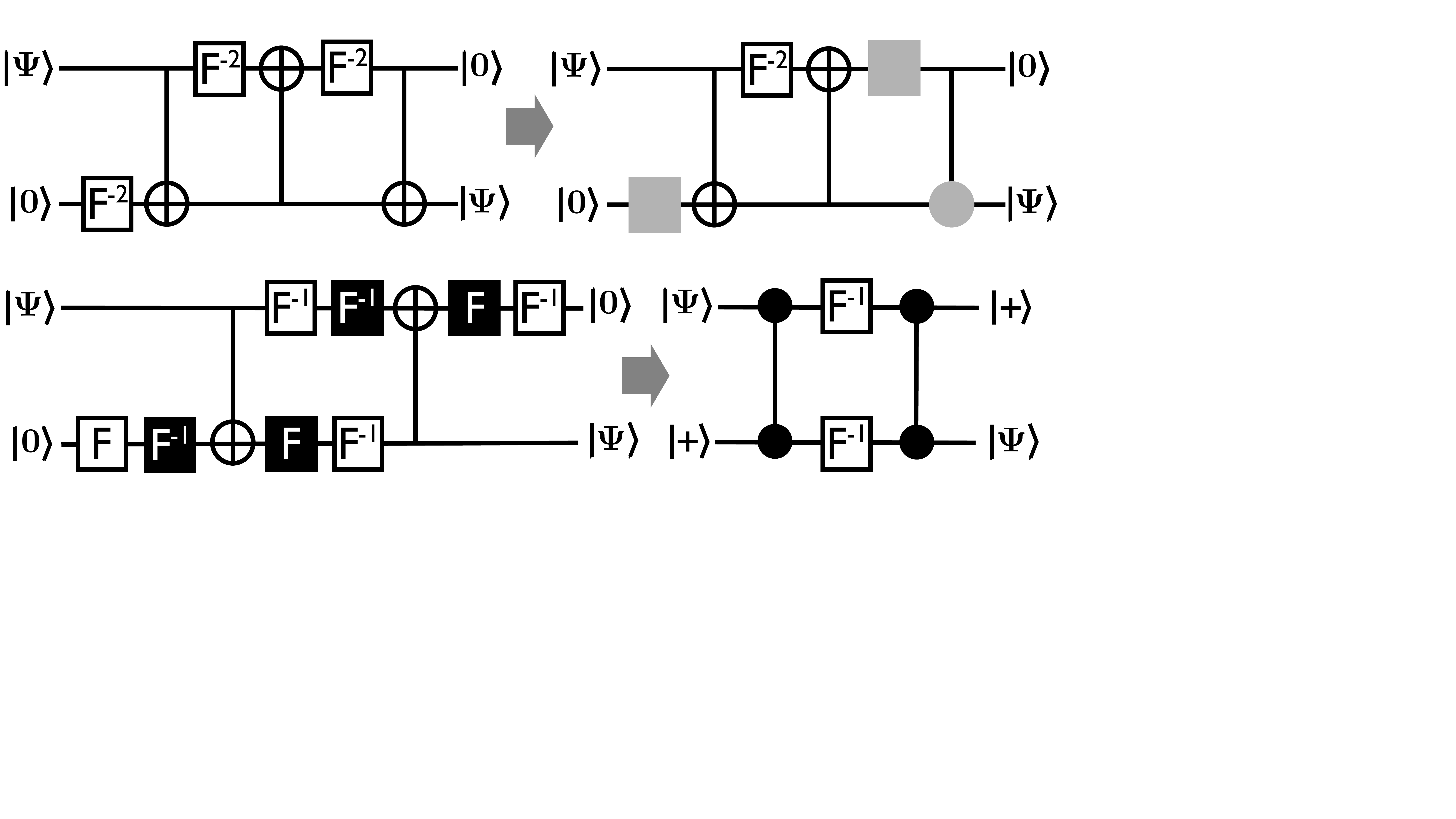}}
\put(-2.2,2.3){(A)}
\put(2.3,2.3){(B)}
\put(-2.2,.4){(C)}
\put(3.2,.4){(D)}
\end{picture}
\end{center}
	\caption{Simplification of the qudit SWAP gate when one of the inputs is the $\ket{0}$ state. (A) Full SWAP from (\ref{Swap1}), (B) shaded gates are redundant for input $\ket{0}_2$, (C) inserting $\hat{F}\hat{F}^{-1}$ on either side of the target on CSUM gates, (D) simplification using (\ref{Sum2}).}
	\label{fig1}
\end{figure}

\begin{figure}[t]
\begin{center}
\setlength{\unitlength}{1cm}
\mbox{
\begin{picture}(15,4.3)
\put(0.25,-1){\includegraphics[scale=0.135]{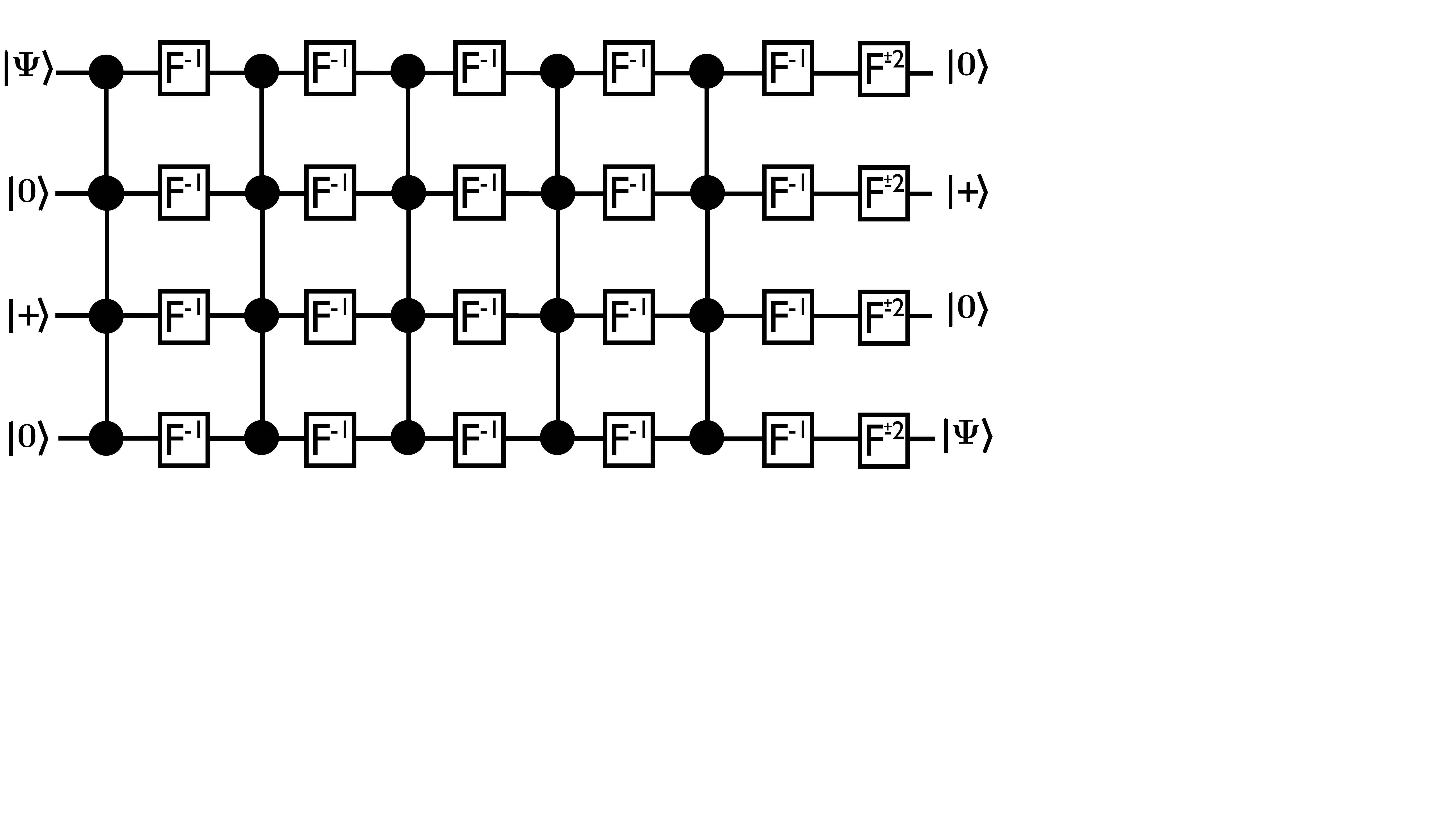}}
\put(0.25,-4.25){\includegraphics[scale=0.125]{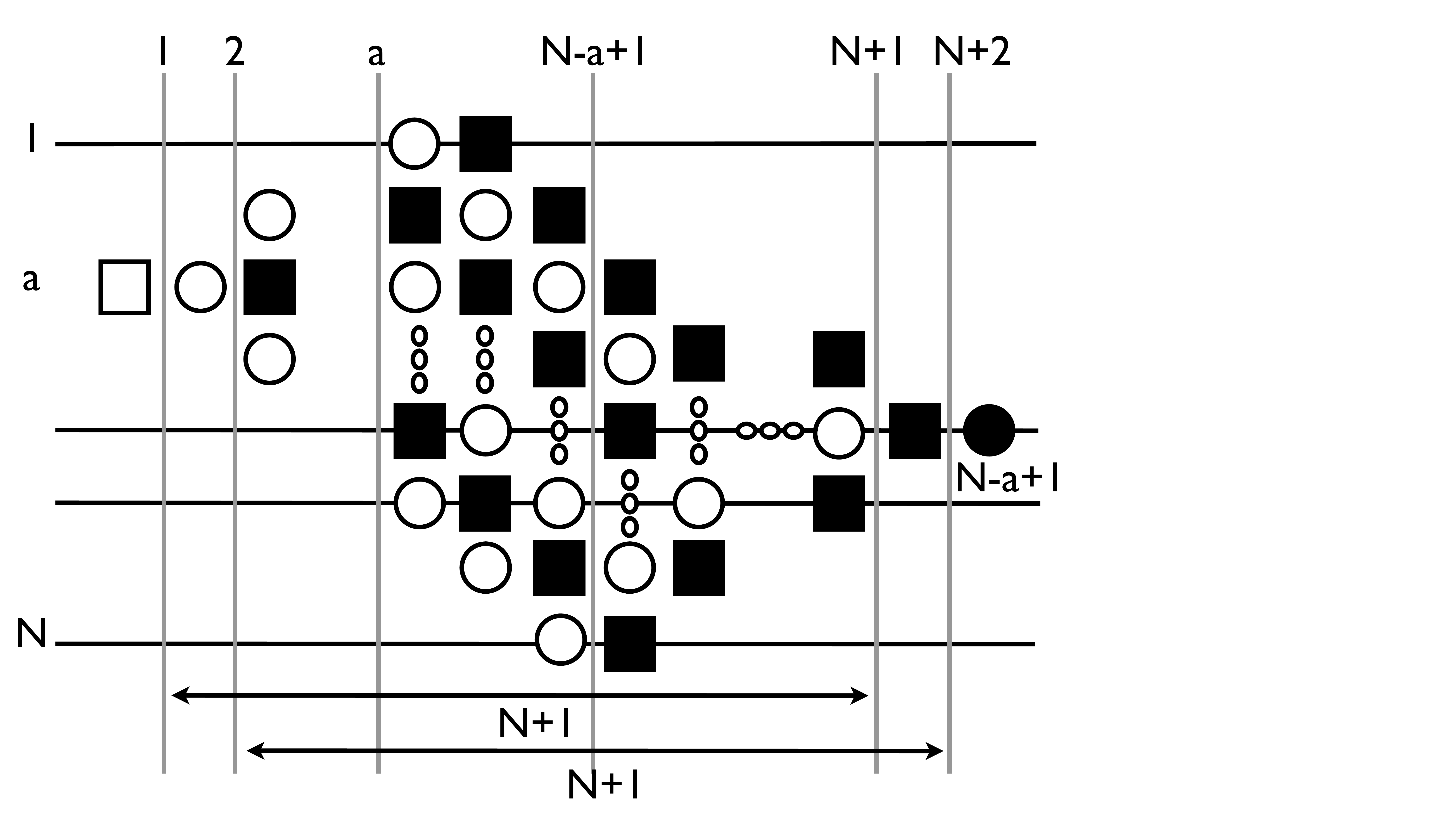} }
\put(0,4.1){\large(A)}
\put(0,0.5){\large(B)}
\end{picture}}
\end{center}
\vspace{4.0cm}
	\caption{(A) Perfect mirror inversion circuit for qudits. Although particular input states are shown, the mirror action is independent of the input. (B) Evolution of the $\hat{Z}^l$ (and $\hat{X}^l$) under repeated actions of $\hat{F}^{-1} S$. The horizontal axis gives the number applications of $\hat{F}^{-1} S$, while the vertical axis is the qubit position. The $\hat{Z}^l$ ($\Box$) and $\hat{X}^l$ ($\bigcirc$) operators evolve according to (\ref{Eqs:BasicAlgRels}), (black shape indicating a negative power of the correponding operator). After $N+1$ repetitions the $\hat{Z}^l$ operator has moved to its mirror spatial position along the chain but has suffered $l\rightarrow -l$. Negative exponent is corrected by the final $\hat{F}^{-2}$ operation.}
	\label{Fig2}
	\vspace{-0.5cm}
\end{figure}

Fig. \ref{fig1} shows the simplified qudit SWAP gate when one of the inputs is $\ket{0}$. It builds an $N-$qudit perfect mirror circuit [Fig \ref{Fig2}(A)] by replicating the simplified two qudit SWAP gate in Fig.\ref{fig1}(D) $N-1$ times and adding redundant  CPHASE and $\hat{F}$ gates, where $\ket{+} = \hat{F}^{+1} \ket{0}$. Through the repeated applications of $\overline{S}\overline{F}$, with the {\em global qudit gates} $\overline{F}\equiv\prod_{j=1}^N\,\hat{F}_{(j)}$, and $\overline{S}\equiv \prod_{j=1}^{N-1}\hat{S}_{(j,j+1)}$, one can transport the unknown quantum state  from one end of the qudit chain to the other~\footnote{Although the circuit thus obtained  does not need the final sequence $\overline{S} \hat{F}^{-1} \hat{F}^{\pm 2}$, this extra segment is required to mirror invert for arbitrary inputs.}. As it stands, it would seem from Fig. \ref{Fig2}(A) that perfect mirror inversion requires the particular initial state $|\Psi\rangle\otimes \ket{0}\otimes\ket{+}\cdots\otimes\ket{0}$. We now show that this is not the case.

It was shown for the qubit case that the transport does not depend on the input states \cite{Fitzsimons:2006p852}. The proof relies on showing that $
(\overline{H}\,\overline{S})^{N+1} \hat{X}^l_{(a)} = \hat{X}^l_{(N+1-a)}(\overline{H}\,\overline{S})^{N+1}$ and
$(\overline{H}\,\overline{S})^{N+1} \hat{Z}^l_{(a)} = \hat{Z}^l_{(N+1-a)}(\overline{H}\,\overline{S})^{N+1}$ which implies that any $N-$qubit density matrix is spatially inverted after the action of $(\overline{H}\,\overline{S})^{N+1}$. For qudits, it suffices to show the analogous relations, i.e.
\begin{subequations}
\label{relation}
\beqy
\overline{F}^{\pm2}(\overline{F}^{-1}\overline{S})^{N+1} \hat{X}^l_{(a)} &=& \hat{X}^l_{(N+1-a)}\overline{F}^{\pm2}(\overline{F}^{-1}\overline{S})^{N+1}, \label{relax} \\  
\overline{F}^{\pm2}(\overline{F}^{-1}\overline{S})^{N+1} \hat{Z}^l_{(a)} &=& \hat{Z}^l_{(N+1-a)}\overline{F}^{\pm2}(\overline{F}^{-1}\overline{S})^{N+1}. \label{relaz}
\eneqy
\end{subequations}
We now develop the basic algebraic relations used to prove (\ref{relation}). These relations will be very important in a generalization from qudits to continuous variables, as showing that they also hold for the CV case immediately verifies the validity of the mirror circuit in that case. To simplify the notation, the overbar denoting the global pulse is dropped unless stated otherwise by a subscript i.e. $\overline{F}\rightarrow \hat{F}$, but $\hat{F}_{(a)}\rightarrow \hat{F}_{(a)}$. Direct calculation shows
\beq 
\hat{F}_{(a)}^{\pm 2} \hat{Z}_{(a)}^l \hat{F}_{(a)}^{\pm 2} = \hat{Z}_{(a)}^{-l}\;\;,\label{minus}
\eneq
and $\hat{F}_{(a)} \hat{X}_{(a)}^l = \hat{Z}_{(a)}^l \hat{F}_{(a)}$. The operator (\ref{Phase1}) can be written as $\hat{S}_{(m,n)} = \sum_{ij} \zeta^{-ij} \hat{Z}^i_{(m)} \otimes \hat{Z}^j_{(n)}$, to obtain 
\begin{subequations}
\label{Eqs:BasicAlgRels}
\beqy
\hat{S}_{(a,a+1)}  \hat{X}^l_{(a)} &=& \hat{X}^l_{(a)} \hat{Z}^l_{(a+1)} \hat{S}_{(a,a+1)}\;\;, \label{13}\\
\hat{S}_{(a,a+1)}  \hat{X}^l_{(a+1)} &=& \hat{X}^l_{(a+1)} \hat{Z}^l_{(a)} \hat{S}_{(a,a+1)} \;\;,\\
\hat{S}_{(a,a+1)}  \hat{Z}^l_{(a)} &=& \hat{Z}^l_{(a)} \hat{S}_{(a,a+1)}\;\;,\\
\label{trX}
(\hat{F}^{-1}\hat{S})  \hat{X}^l_{(a)} &=& \hat{X}^l_{(a-1)} \hat{Z}^{-l}_{(a)} \hat{X}^l_{(a+1)} (\hat{F}^{-1}\hat{S})\;\;, \\
\label{trZ}
(\hat{F}^{-1}\hat{S})  \hat{Z}^l_{(a)} &=& \hat{X}^l_{(a)} (\hat{F}^{-1}\hat{S})\;\;.\label{17}
\eneqy
\end{subequations}
\hspace{0.01cm}
\hspace{0.1cm}
To derive Eqs.~\eqref{relation} a graphical method~\cite{Fitzsimons:2006p852} is used [Fig. \ref{Fig2}(B)] to obtain $(\hat{F}^{-1} \hat{S})^{N+1} \hat{X}^{l}_{(a)}   = \hat{X}^{-l}_{(N+1-a)}(\hat{F}^{-1} \hat{S})^{N+1}$ and $(\hat{F}^{-1} \hat{S})^{N+1} \hat{Z}^{l}_{(a)} = \hat{Z}^{-l}_{(N+1-a)}(\hat{F}^{-1} \hat{S})^{N+1}$. It follows [via (\ref{minus})] that extra final $\hat{F}^{\pm 2}$ corrects the exponent $l\rightarrow -l$. Fig. \ref{Fig2}(B) shows the evolution of an initial $\hat{Z}^l$ operator after the consecutive action of $\hat{F}^{-1} S$. It is extended to $N+2$ so the evolution of both $\hat{X}^l$ and $\hat{Z}^l$ can be fully appreciated. It is important to note that the property $\hat{Z}^d = \hat{\mathbb{I}} = \hat{X}^d$ was never used, thus allowing us to use the same method for the CV case which lacks this cyclic property.

{\em Continuous variable mirror transport:-} For continuous variables \cite{Braunstein:2005p3364} the same argument is used. The CV analogues of the generalised Pauli operators are $\hat{X}(q)\equiv \zeta^{-q\hat{p}}$, $\hat{Z}(p)\equiv \zeta^{p\hat{x}}$ where $\zeta = e^{i/\hbar}$, are non-commutative, $\hat{X}(q)\hat{Z}(p)=\zeta^{-qp}\hat{Z}(p)\hat{X}(q)$, and have the following action on the computational basis (position eigenstates $\hat{x}\ket{q}=q\ket{q}$): $\hat{X}(q) \ket{s}= \ket{s+q}$, $\hat{Z}(p) \ket{s} = \eta^{sp} \ket{s}$. By combining $\hat{X}(q)$ and $\hat{Z}(p)$, the displacement operator $\hat{D}(\alpha)=\exp(\alpha \hat{a}^\dagger-\alpha^*\hat{a})$ can be formed. Given that any valid infinite dimensional density operator can be decomposed into a sum over coherent states via the $P-$function \cite{Hillery:1984p5397}, $\hat{\rho}=\int\,d^2\alpha P(\alpha)\ket{\alpha}\bra{\alpha}$, and that $\ket{\alpha}=\hat{D}(\alpha)\ket{0}$, the operators $\hat{\Xi}(q,p)=\hat{X}(q)\hat{Z}(p);\;{q,p}\in\mathbb{R}^2$ form a basis for $\mathcal{H}^\infty$. This shows that CV-mirror transport protocol can be used to mirror {\em any} initial state of the CV-chain.  

Following the literature on CV-gates \cite{Braunstein:2005p3364}, basic gates are defined using the same notation as in qudit case. The position representation will be used for all operators. The Fourier transformation is $\hat{F}(\sigma) \ket{x} = \left(\sigma\sqrt{\pi}\right)^{-1} \int \textrm{d}y\,  \zeta^{ 2 xy /\sigma^2} \ket{y}$, satisfying $\hat{F}^4 = \hat{\mathbb{I}}$ and $\hat{F}^2 \ket{x} = \ket{-x}$ upon setting $\sigma=\sqrt{2}$. The basic gates needed are
\begin{subequations}
\begin{eqnarray}
\hat{U}^\pm_{SUM ({12})} = \int \textrm{d}x\,\ket{x}\bra{x}_{(1)} \otimes \hat{X}_{(2)}(\pm x)=\widehat{CN}^\pm_{(12)}&& \\
\label{phaseZZ}
\hat{U}^{\pm}_{CPHASE} = \int \textrm{d}x\,\ket{x}\bra{x}_{(1)} \otimes \hat{Z}_{(2)}(\pm x) = \hat{S}^{\pm}_{(1,2)}&& 
\end{eqnarray}
\end{subequations}
where the identity $\hat{U}^{\pm}_{CPHASE} = \hat{F}_{(2)} \widehat{CN}^\pm_{(12)} \hat{F}^{-1}_{(2)}$ holds again. The mirror circuit can then be built by using the $+$ version of the defined gates and taking $\sigma = \sqrt{2}$.
 
Note that using the $+$ version of these gates allows to define the circuit independent of the basis~\cite{Wang:2001p3095}. With this in mind the reader shouldn't be surprised that the relations
\begin{eqnarray}
&&\!\!\!\!\!\!\!\!\!\!\!\!\hat{F}^{\pm 2} \hat{Z}(s) \hat{F}^{\pm 2} = \hat{Z}(-s),\; \hat{F} \hat{Z}(s) = \hat{X}(-s) \hat{F},\; \hat{F} \hat{X}(s) = \hat{Z}(s) \hat{F}\nonumber \\
&&\!\!\!\!\!\!\!\!\!\!\!\!\hat{F}^{-1} \hat{X}(s) =\hat{ Z}(-s) \hat{F}^{-1},\; \hat{F}^{-1} \hat{Z}(s) = \hat{X}(s) \hat{F}^{-1}  
\end{eqnarray}
are also valid in this CV case. Since the algebraic relations~\eqref{trX},~\eqref{trZ} are the same, the same proof used for the qudit case is valid in the CV case. Alternatively, we can use the traditional representations of SUM and Fourier transform in CV \cite{Wang:2001p3095}, i.e. without using the representation~\eqref{phaseZZ}, where the SUM gate can be represented as 
$\hat{U}_{SUM(12)}=\widehat{CN}_{(12)}=\exp \left( -i\hat{q}_1\otimes \hat{p}_2\right)$, which has the following actions:
\begin{subequations}
\beqy
\widehat{CN}_ {({12})}: \hat{X}_1(x) \otimes 1_2 \rightarrow  \hat{X}_1(x)\otimes \hat{X}_2(x)\;\;,&&\\
\hat{Z}_1(p) \rightarrow  \hat{Z}_1(p),\;\;\;
 \hat{X}_2(x) \rightarrow  \hat{X}_2(x)\;\;,&&\\
1_i \otimes \hat{Z}_j(p) \rightarrow  \hat{Z}^{-1}_i(p)\otimes \hat{Z}_j(p)\;\;,&&\\
\hat{F}: \hat{X}\rightarrow\hat{Z},\;\;\;\; \hat{Z}\rightarrow \hat{X}^{-1}\;\;.&&
\eneqy
\end{subequations}
With these and $\overline{F}\equiv \prod_{j=1}^N \hat{F}_{(j)}$, and $\overline{S}=\prod_{j=1}^{N-1}\hat{S}_{(j,j+1)}$, we arrive at relations analogous to \eqref{trX} and \eqref{trZ} $\overline{F}^{-1}\overline{S}:  \hat{X}_{(a)} \rightarrow \hat{X}_{(a-1)} \hat{Z}^{-1}_{(a)} \hat{X}_{(a+1)}$ and $\hat{Z}_{(a)} \rightarrow \hat{X}_{(a)}$, so
\beqy
\nonumber \overline{F}^2(\overline{F}^{-1} \overline{S})^{N+1} :\hat{X}(q)_{(a)}   &\rightarrow& \hat{X}(q)_{(N+1-a)}\\
 \hat{Z}(p)_{(a)}   &\rightarrow& \hat{Z}(p)_{(N+1-a)}\label{CVcase}
\eneqy
as desired. Thus (\ref{CVcase}) proves that the circuit shown in Fig. \ref{Fig2}(A), with the analogous CV gates performs perfect quantum mirroring on a chain of CV systems. 

{\em Circuit-QED quantum mirroring:-} The above CV quantum mirror transport can be realised in Circuit-QED \cite{Blais:2004p3701}. The goal is to spatially mirror quantum state of $N-$CV modes held in chain of superconducting coplanar waveguides (CPWs), nearest-neighbour-coupled by switchable Cooper Pair Boxes (CPBs). To this end, $\overline{F}$ and $\overline{S}$ ought to be executed, where for $\overline{S}$, transformation $\exp \left( -i\hat{q}_1\otimes \hat{q}_2\right)$ needs to be generated. We consider the arrangement where the CPB coupling adjacent 1/4-wave CPWs can be switched between being biased by the difference or the sum of the voltages of the two CV modes. 

The Hamiltonian and the resulting master equation for the two CV modes (operators $\hat{a}, \hat{b}$) and CPB ($\sigma_{\pm}$ operators) operating at the charge degeneracy point are  \cite{Blais:2004p3701}
\begin{subequations}
\label{Eq:Hmasterinit}
\begin{eqnarray}
&&\!\!\!\!\!\!\!\!\!\!\!\!\hat{H}_{\pm}/\hbar=\omega_a \hat{a}^\dagger \hat{a}+ \omega_b \hat{b}^\dagger \hat{b}\nonumber\\
&&\!\!\!\!\!\!+\frac{\omega_0}{2}\hat{\sigma}_z-\left[g_a(\hat{a}+\hat{a}^\dagger)\pm g_b(\hat{b}+\hat{b}^\dagger)\right]
(\hat{\sigma}_++\hat{\sigma}_-),\label{circuit-qed}\\
&&\!\!\!\!\!\!\!\!\!\!\!\!\dot{\tilde{\rho}}=\frac{i}{\hbar}[\hat{H}_\pm,\tilde{\rho}]+\left(\frac{\gamma}{2}{\mathcal L}(\hat{\sigma}_-)+\frac{\kappa_a}{2}{\mathcal L}(\hat{a})+\frac{\kappa_b}{2}{\mathcal L}(\hat{b})\right)\tilde{\rho}\label{big-master}
\eneqy
\end{subequations}
where $\hat{H}_{\pm}$ is the dynamics for the case of sum or difference bias conditions, $\omega_{a/b}$ are the mode frequencies of the resonators, $g_{a/b}$ are the couplings between the CPB and each CPW, $\omega_0$ is the CPB splitting, $\gamma, \kappa_{a/b}$ are the decay rates for the CPB and two CPWs, and ${\mathcal L}(\hat{A})\tilde{\rho}\equiv 2\hat{A}\tilde{\rho}\hat{A}^\dagger-\{\hat{A}^\dagger \hat{A},\tilde{\rho}\}$. 

\begin{figure}[t]
\begin{center}
\setlength{\unitlength}{1cm}
\begin{picture}(4,3.4)
\put(-3,-1.3){\includegraphics[width=9cm,height=6cm]{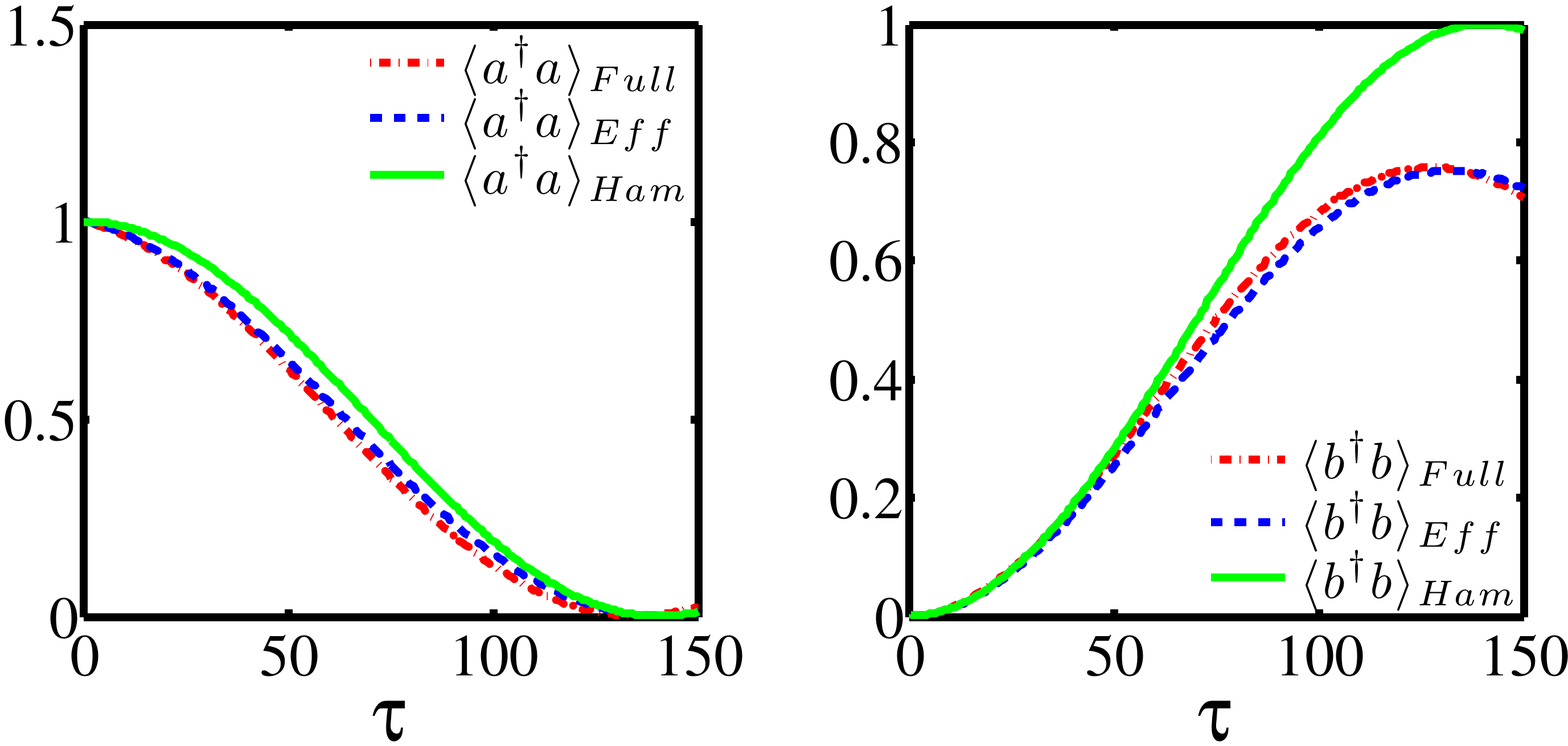}}
\end{picture}
\end{center}
	\caption{Quantum evolution of the two CPW modes in the adiabatic limit for $\hat{H}_-$, taking $|\psi(\tau=0)\rangle=|\alpha=1,\beta=0,1\rangle$, (the two modes in coherent states $|\alpha\rangle\otimes|\beta\rangle$ and CPB state is $|1\rangle$), and taking $\omega_0=15$GHz, $\omega_{a/b}=3$GHz, $g_{a/b}=200$MHz, $\gamma=15$MHz, $\kappa_{a/b}=1$MHz. Plots show the reduced mode dynamics of the full master equation (\ref{big-master}) [{\em Full}], the effective master equation (\ref{master-effective}) [{\em Eff}], and subject only to the effective Hamiltonian (\ref{Heff}) [{\em Ham}], with $\tau=\omega_a t/2\pi$, the nominal oscillation period of the CPW modes. }
	\label{following}
	\vspace{-0.5cm}
\end{figure}

Assuming that CPB dynamics dominate all other time scales in (\ref{Eq:Hmasterinit}), i.e. $\omega_0>\omega_{a/b}, g_{a/b}$, and $\gamma> \kappa_{a/b}$, the fast dynamics of the CPB can be eliminated~\cite{WGardiner:1983p5596}. The equation for $\rho = \textrm{Tr}_{CPB}\{ \tilde{\rho} \}$, describing dynamics of the two coupled CV modes is (for $g_a=g_b$)
\begin{subequations}
\label{Eq:Hmasterfin}
\beqy
&&\!\!\!\!\!\!\dot{\rho} = -i/\hbar[\hat{H}_{eff},\rho]+\sum_{j=\hat{a},\hat{b}}\kappa_j {\mathcal L}(j)\rho+\eta{\mathcal L}(\hat{s})\rho\label{master-effective}\;\;, \\
&&\!\!\!\!\!\!\hat{H}_{eff} = \hbar\omega_a\hat{a}^\dagger \hat{a} + \hbar\omega_b \hat{b}^\dagger \hat{b} + \hbar\chi \hat{s}^2\;\;,\label{Heff}
\eneqy
\end{subequations}
where $\hat{s}=\hat{X}_a\pm \hat{X}_b$ and $\hat{X}_{a/b}$ are the position operators of the two CV modes $\chi=g^2\omega_0/[(\gamma/2)^2+\omega_0^2]$, and $\eta=(\gamma/2)/[(\gamma/2)^2+\omega_0^2]$. Fig. \ref{following} shows that this approximation is valid for several hundred oscillator periods.

\begin{figure}[t]
\setlength{\unitlength}{1cm}
\begin{center}
\begin{picture}(4.5,6.4)
\put(-2.95,-1.6){\includegraphics[scale=0.455]{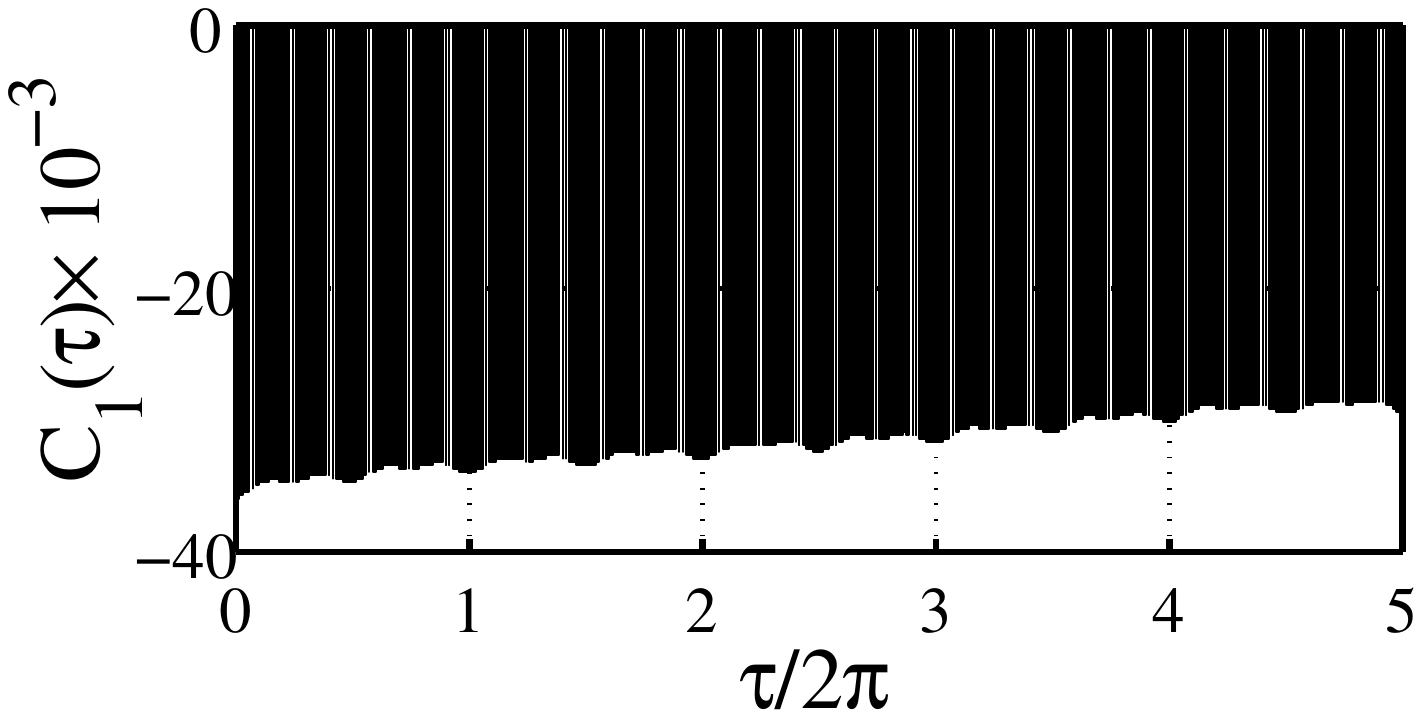}}
\put(-2.95,-5.3){\includegraphics[scale=0.455]{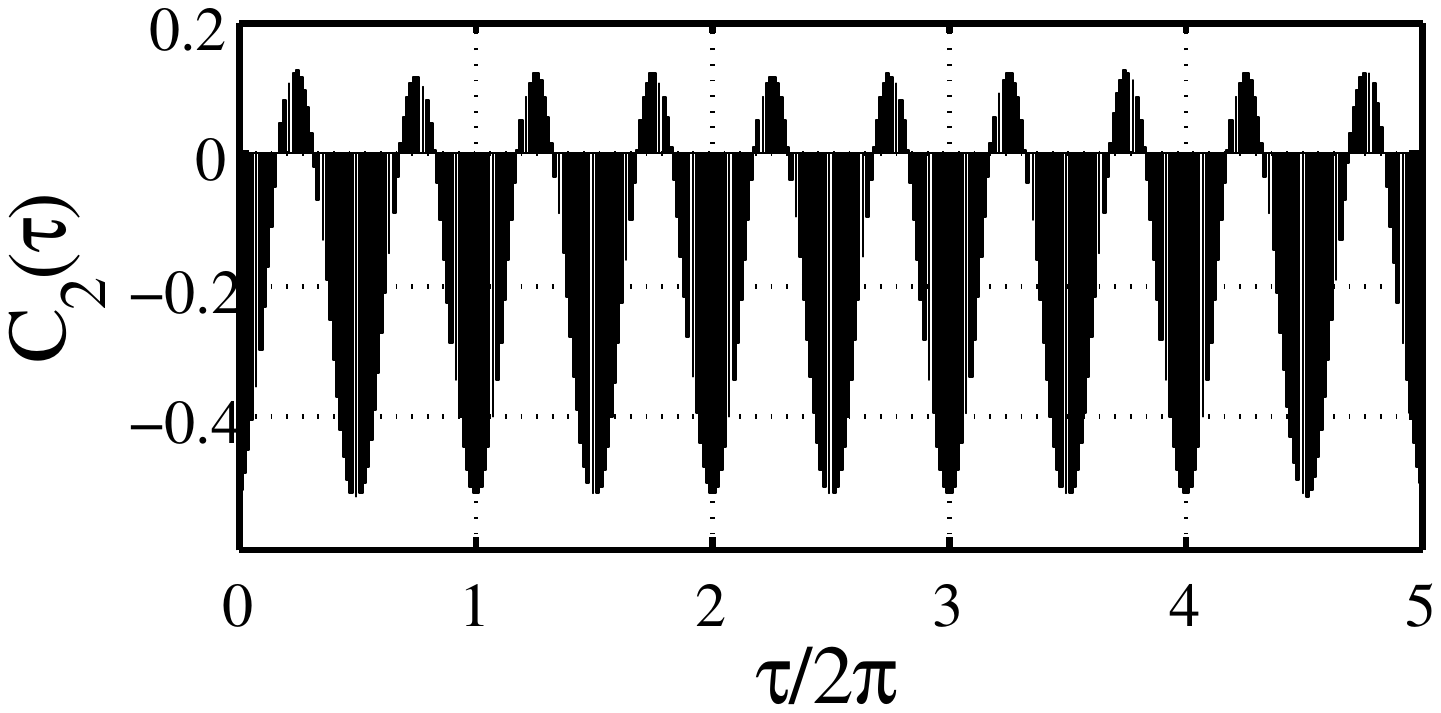}}
\end{picture}
\end{center}
	\caption{Pulse trains to effect $\hat{\bar{U}}^\prime_a\equiv \exp(i\hat{X}_{a^\prime}^2/10)$ for parameters of Fig. \ref{following} to an accuracy of ${\cal F}=\sqrt{|{\rm Tr}(\hat{U}^\dagger \hat{\bar{U}}^\prime_a)|}/\sqrt{{\rm Tr}(\hat{\bar{U}}^{\prime\,\dagger}_a\hat{\bar{U}}^\prime_a)}$ = 99.9\% \cite{Khaneja:2005p5829}. Total time to effect desired unitary operator $U^\prime_a$ is thus 50 oscillator cycles.}
	\label{controls}
	\vspace{-0.5cm}
\end{figure}

To generate the unitary operator $\hat{U}_{PHASE(12)}$, consider the effective Hamiltonian (\ref{Heff}), with the mode frequencies and effective coupling strengths time dependent $\omega_a(t)=\omega_b(t)$. Such rapid tuning of the resonator frequencies is achievable \cite{Sandberg}. The canonical transformation $\hat{H}^\prime(t)_{eff}=\hat{T}\,\hat{H}_{eff}(t)\hat{T}^\dagger$ generated by $\hat{T}(a,b)\equiv\exp(-\pi [\hat{a}^\dagger \hat{b}-\hat{b}^\dagger \hat{a}]/4)$ decouples (\ref{Heff}) into two separate harmonic systems. The resulting Hamiltonian is $\hat{H}_{eff}^\prime (t)=\omega_+(t)(\hat{a}^{\prime\,\dagger}\hat{a}^\prime+\hat{b}^{\prime\,\dagger}\hat{b}^\prime)+2\chi(t)\hat{X}^2_{a^\prime/b^\prime}$, where $\omega_+(t)=\omega_a(t)$ with either $\hat{X}^2_{a^\prime}$ for sum voltage biasing in (\ref{big-master}) or $\hat{X}^2_{b^\prime}$ for difference voltage biasing. Under this canonical transformation the operator $\hat{U}_{CPHASE}$ also separates $\hat{U}_{CPHASE}^\prime=\exp(i\hat{X}_{a^\prime}^2)\exp(-i\hat{X}_{b^\prime}^2)$. Further, the Fourier gates applied homogeneously to both modes $\hat{a}$ and $\hat{b}$ transform to homogeneously applied Fourier gates on the modes $\hat{a}^\prime$ and $\hat{b}^\prime$. This gate is achieved via the natural evolution of the uncoupled CPW resonators. We describe how to generate the $\hat{U}_{CPHASE}^\prime$ by showing how to generate the first component $\hat{U}^\prime_a=\exp(i\hat{X}_{a^\prime}^2)$: $(i)$ numerically find time dependencies for  $\omega_+(t)$ and $\chi(t)$,  resulting in generation of $\hat{U}^\prime_a$ while simultaneously $\hat{b}^\prime$ acquires a known rotation $\exp(-i\theta \hat{b}^{\prime\,\dagger}\hat{b}^\prime)$; $(ii)$ Switch to difference voltage biasing, evolving via a similar set of pulses, to effect $\hat{U}^\prime_b$, while $\hat{a}^\prime$ acquires an identical known rotation. These extra rotations are then absorbed into the preceding and succeeding Fourier gates in the overall application of the Fig.~\ref{Fig2}(A). To derive the required time dependent pulse we focus on the $\hat{a}^\prime$ dynamics in $\hat{H}_{eff}^\prime$ and move to a new time variable $\tau=\bar{\omega} t$, where $\omega_+(t)=\bar{\omega}(1+C_1(t)/50)$, $\chi(t)=\bar{\chi}(1+C_2(t)/50)$, and write $\hat{\bar{H}}^\prime_{a}(\tau)=[1+C_1(\tau)/50] \hat{a}^{\prime\,\dagger}\hat{a}^\prime+[\epsilon+C_2(\tau)/50]\hat{X}^2_{a^\prime}$. $C_1(\tau)$ and $C_2(\tau)$ are control functions and $\epsilon=2\bar{\chi}/\bar{\omega}\sim 1\times 10^{-3}$ for the parameters of Fig. \ref{following}.  To obtain pulse sequences which attain 99.9\% accuracy in simulating $U^\prime_a$, we use GRAPE-based numerical optimization algorithms \cite{Khaneja:2005p5829}  in 50 cycles of the oscillator and with a Fock state truncation of 70 \footnote{Matlab code to show pulse and degree of fidelity to target unitary operator can be found on http://www.quantumscience.info/node/346.}. The pulses, displayed in Fig. \ref{controls}, require tuning the CPW resonators by 2.4MHz and the CPB by 150MHz.  It is clear that if one can generate larger detunings of the oscillators and CPB, the time required to accurately simulate $U^\prime_a$ can be even shorter, thus allowing several mirror iterates to be executed before the resonator decay degrades the dynamics.

In summary a protocol for perfect quantum mirroring of a chain of coupled qudits or CV systems was proposed. Such capability will allow the quantum transport of entire entangled qudit or CV quantum states via quantum mirroring. The protocol uses only global applications of higher dimensional versions of the Hadamard and CPHASE gates. We anticipate that  CV universal quantum computation can be achieved if the protocol is augumented with the capability of performing a conditional phase gate between the end two CV qubits in the chain. This would require a large cross-Kerr nonlinear interaction between these CV systems. Finally, it was shown how to implement CV quantum mirroring in Circuit-QED. 

We thank European Commission FP6 IST FET QIPC project QAP Contract No. 015848, DEST ISL Grant CG090188 (SR \& JT) and Mazda Foundation for Arts and Science (GAPS).

\end{document}